\documentclass[a4paper,11pt]{article}
\usepackage{jcappub} 
\usepackage{float}
\usepackage{cancel}
\usepackage{upgreek}
\usepackage{siunitx}
\usepackage{placeins}
\usepackage{orcidlink}
\usepackage{comment}
\usepackage[normalem]{ulem}
\usepackage{booktabs}
\usepackage{appendix}
\usepackage{indentfirst}
\usepackage{xcolor}
\usepackage{amsmath,mathtools,amsfonts,amsthm,amssymb,amstext}
\usepackage[T1]{fontenc} 
\usepackage{hyperref}
\usepackage{tikz}
\usepackage{booktabs}

\graphicspath{{figures/}}
\hypersetup{
  colorlinks   = true, 
  urlcolor     = blue, 
  linkcolor    = blue, 
  citecolor   = blue 
}
\usepackage{color}
\usepackage[dvipsnames]{xcolor}

\newcommand{\mpl}{m_P}
\newcommand{\efold}{\textit{e}-fold}
\newcommand{\efolds}{\textit{e}-folds}
\newcommand{\rhog}{\rho_{\nabla}}
\newcommand{\rhok}{\rho_{K}}
\newcommand{\rhodk}{\rho_{\delta K}}
\newcommand{\rhodko}{\rho_{\delta K,0}}
\newcommand{\rhob}{\overline{\rho}}
\newcommand{\rhop}{\rho_{\delta\phi}}
\newcommand{\rhogo}{\rho_{\nabla,0}}

\newcommand{\rhobo}{\overline{\rho}_{0}}

\newcommand{\grchombo}{\texttt{GRChombo}}
\newcommand{\kcl}{Theoretical Particle Physics and Cosmology Group, Physics Department, King's College London, Strand, London WC2R 2LS, United Kingdom}

\begin{document}
{\hfill KCL-PH-TH/2025-33 \vspace{-15pt}}
\title{Non-linear Dynamics and Primordial Black Hole Formation During Kination}
 \author[a]{Cheng Cheng,\orcidlink{0009-0009-9070-870X}}
 \emailAdd{cheng.2.cheng@kcl.ac.uk}
 \author[a]{Panagiotis Giannadakis,\orcidlink{0009-0007-5763-9576}}
 \emailAdd{panagiotis.giannadakis@kcl.ac.uk}
 \author[a]{Lucien Heurtier,\orcidlink{0000-0003-3153-7225}}
 \emailAdd{lucien.heurtier@kcl.ac.uk}
 \author[a]{Eugene A. Lim,\orcidlink{0000-0002-6227-9540}}
 \emailAdd{eugene.a.lim@gmail.com}
 \affiliation[a]{\kcl}

\abstract{
  We investigate the effects of large scalar inhomogeneities during the kination epoch, a period in which the universe's dynamics are dominated by the kinetic energy of a scalar field, by fully evolving the Einstein equations using numerical relativity. By tracking the non-linear growth of scalar perturbations with both sub-horizon and super-horizon initial wavelengths, we are able to compare their evolution to perturbative results. Our key findings show that in the deep sub-horizon limit, the perturbative behaviour remains valid, whereas in the super-horizon regime, non-linear dynamics exhibit a much richer phenomenology. Finally, we discuss the possibility of primordial black hole formation from the collapse of such perturbations and assess whether this process could serve as a viable mechanism to reheat the universe in the post-inflationary era.}
\maketitle
  
\section{Introduction}

The standard cosmological narrative begins with a period of slow-roll inflation~\cite{Guth:1980zm, Linde:1981mu, Albrecht:1982wi}, during which the dynamics are governed by a scalar field—the inflaton—rolling down a nearly flat potential. This phase provides solutions to the horizon and flatness problems while stretching quantum fluctuations of the inflaton field that later seed the formation of large-scale structures. After inflation ends, it is typically assumed that the inflationary sector reheats the universe by transferring most of its energy to Standard Model particles. This marks the beginning of the Hot Big Bang phase, initially characterized by a radiation-dominated plasma with equation-of-state parameter $w = 1/3$, which eventually transitions to matter domination with $w = 0$ as the universe cools. 

Interestingly, there are no strong observational constraints on the equation of state of the universe during the period between the end of inflation and the onset of reheating. This opens up the possibility for a variety of cosmological scenarios. One such possibility is that the post-inflationary--but pre-reheating--epoch is dominated by the kinetic energy of a rolling scalar field, which may either be the inflaton or another scalar degree of freedom. This phase, known as \textit{kination}, was first proposed in~\cite{Joyce_1997}, characterised by equation-of-state parameter $w = 1$. For a more comprehensive review of kination, see~\cite{gouttenoire2022kinationcosmologyscalarfields}.

Kination is especially relevant in the context of string theory, where ultra-light moduli fields can dominate early-universe dynamics~\cite{cicoli2023stringcosmologyearlyuniverse}. String cosmology models involving either the inflaton or other moduli may naturally give rise to kination if the relevant modulus potential features a steep region at the end of inflation--where the potential decreases exponentially, enabling the field to acquire significant kinetic energy--followed by a flat region where the field eventually slows down. What happens afterward depends on the exact shape of the potential.

If the potential exhibits runaway behaviour, the field slows down due to the Hubble friction and may even initiate a secondary period of slow-roll inflation. This feature is common in \textit{quintessential} inflation models~\cite{Peebles:1998qn,Neupane:2007jm, de_Haro_2021, Bettoni_2022, Dimopoulos:2017zvq, Dimopoulos:2001ix} and more generally, in non-oscillatory inflation scenarios~\cite{Ellis:2020krl,Heurtier:2022rhf}, where the inflaton field undergoes $(i)$ a slow-roll phase in the early universe, $(ii)$ a phase of acceleration as the potential decreases sharply and the universe enters a kination era, and $(iii)$ a second phase of slow-roll during which the field decelerates to eventually either behave as an effective cosmological constant or track the universe's background energy density at late times~\cite{Zlatev:1998tr,Steinhardt:1999nw,Ellis:2020krl,Heurtier:2022rhf, apers2024stringtheoryhalfuniverse}. Alternatively, if the potential features a local minimum, the field may eventually settle into it, acquiring a finite mass and oscillating before decaying to reheat the universe, as occurs in Large Volume Scenarios (LVS)~\cite{Balasubramanian_2005,Conlon_2005, apers2024stringtheoryhalfuniverse}. 

The existence of a long kination era within the cosmological timeline may leave imprints in cosmological data that could be searched for in the foreseeable future~\cite{Ghoshal:2022ruy, Gouttenoire:2021jhk, Soman:2024zor}. However, the authors of~\cite{eröncel2025universalbounddurationkination, apers2024stringtheoryhalfuniverse, mosny2025selftrackingsolutionsasymptoticscalar} pointed out that, according to perturbation theory, such a kination era is expected to be unstable due to the growth of scalar-field perturbations. Unlike the homogeneous background field, these perturbations behave like a relativistic perfect fluid with an equation-of-state parameter of $w=1/3$. The authors of~\cite{mosny2025selftrackingsolutionsasymptoticscalar} showed that such scalar perturbations may act as a background for the homogeneous scalar field, which ends up tracking its own perturbation and eventually behaves like radiation. Generally speaking, such considerations suggest that a kination era may not exceed $\sim10 - 11$ \efolds{}  ~\cite{eröncel2025universalbounddurationkination,apers2024stringtheoryhalfuniverse}.

In~\cite{apers2024stringtheoryhalfuniverse, eröncel2025universalbounddurationkination, mosny2025selftrackingsolutionsasymptoticscalar}, the fate of small scalar perturbations is explored in the context of an initially kinetic-energy-dominated background where the gravitational backreaction is neglected. Perturbation theory has provided valuable insights into this regime; however, its main limitation is that it cannot capture strong gravitational effects or non-linear backreaction.

In this work, we instead investigate the evolution of kination in the presence of non-linear perturbations where strong gravitational effects and backreaction can play an important role.
This becomes particularly important during kination, where even initially small scalar field perturbations---especially those of large super-horizon wavelength---can undergo signification amplification. As these modes re-enter the Hubble horizon, they may dominate the local dynamics, inducing sharp \emph{local} gradient inhomogeneities that strongly backreact on the geometry. Consequently, the study of perturbations during kination transitions into a genuine strong-gravity problem. In scenarios where the backreaction becomes sufficiently strong, gravitational collapse may occur, potentially leading to black hole formation. In this context, while perturbation theory provides essential guidance, this regime requires numerical relativity.

We find that for sub-horizon perturbations, even for relatively large initial amplitudes, gravitational backreaction remains weak and does not induce collapse. In this case, scalar perturbations \emph{on average} evolve like radiation for scales much larger than the perturbation scales, in agreement with perturbative predictions. Gravitational backreaction only becomes significant with initial perturbation energy density much larger than the kination background energy density. In this scenario, the notion of a kination background ceases to be well defined. In contrast, for initially linear super-horizon perturbations, gravitational collapse becomes substantially more likely after horizon re-entry. We provide a quantitative analysis of the critical initial overdensity required for collapse and compare it with perturbative expectations. Our results indicate that black hole formation is generally easier during kination than in perturbative analysis. This has important implications for the shape and amplitude of the power spectrum necessary to produce a significant primordial black hole abundance in the early universe.

This paper is organised as follows: In Section~\ref{sect:theory} we provide an introductory framework on kination and our numerical simulation setups, in Section~\ref{sect:results} we present our numerical results for sub-horizon and super-horizon perturbations respectively, focusing on the conditions for black hole formation in the super-horizon case. In Section~\ref{sect:implications} we discuss cosmological implications for the production of primordial black holes in cosmology, and we conclude in Section~\ref{sect: summary}. 

In this work, we use the non-reduced Planck mass, $\mpl^2\equiv\hbar c/G= G^{-1}$ with $\hbar=c=1$ for theoretical analysis, and we use geometric units with $G=c=1$ for simulation works.

\section{Theory and Methodology}\label{sect:theory}
\subsection{Perturbations in Kination Epoch}
In order to study the dynamics during the kination epoch, we consider a single scalar field minimally coupled to gravity with a potential $V(\phi)$ which, in a post-inflationary scenario, is expected to feature a long (trans-Planckian) region where the potential is decaying exponentially such as in LVS models or quintessence. In such potential configurations, the scalar field can roll for an extended period, acquiring large canonical momentum while the potential remains subdominant, namely $V(\phi)\ll \frac{1}{2}\dot{\phi}^2$.
The relevant action is given by:
\begin{equation}
    S=\int d^4x\sqrt{-g}\left(\frac{\mpl^2}{16\pi}R-\frac{1}{2}(\nabla\phi)^2-V(\phi)\right)\,,\hspace{-0.5cm}
\end{equation}
where $R$ denote the Ricci scalar.

For simplicity, we restrict ourselves to the case where the dynamics are dominated by the kinetic energy of the homogeneous scalar field, allowing us to work in the limit where $V(\phi)\rightarrow0$. In this regime, the equation of motion is given by 
\begin{equation}
    \ddot{\phi}+3H\dot\phi-\nabla^2\phi=0\,.
\end{equation}
\begin{equation}
    H^2\equiv\left(\frac{\dot{a}}{a}\right)^2=\frac{8\pi}{3\mpl^2}\left(\frac{1}{2}\dot\phi^2+\frac{1}{2}(\nabla\phi)^2\right)\,.
\end{equation}
Before fully solving Einstein's equations and investigating the effect of non-linear dynamics on the system, let us briefly recap the results from cosmological perturbation theory, in which the full field solution $\phi(t,x)\equiv \bar{\phi}(t) + \delta\phi(t,\vec{x}) $ is decomposed into a homogeneous background solution and small spatial perturbations. The background energy density consists purely of kinetic energy, such that 
\begin{equation}
\rhob\equiv\frac{1}{2}\dot{\bar\phi}^2\sim a^{-6}~.
\end{equation}
Meanwhile, the linear perturbations behave as $\rhop\sim a^{-4}$ in the sub-horizon regime, $k\gg H$, and $\rhop\sim a^{-2}$ in the super-horizon regime, $k\ll H$. For more details on perturbation theory, we refer the reader to Appendix \ref{perturbationtheory}. 

In the homogeneous limit where $\delta\phi = 0$, $\rho\equiv\rhob\sim a^{-6}$, until another fluid with a different equation-of-state parameter starts dominating. However, in the presence of perturbations, the situation becomes more interesting because the energy density of the background field redshifts faster than the energy density of its own perturbations. Whenever a notion of a homogeneous background field can be defined---as is the case in the presence of Gaussian and deep sub-horizon perturbations-- the metric evolution is still well described by standard FLRW cosmology. This could lead to the domination of the perturbations as a new cosmological background radiation fluid, and possibly to tracker solutions in the presence of a subdominant potential, as shown in~\cite{mosny2025selftrackingsolutionsasymptoticscalar}. 

However, for perturbations with sizes comparable to or larger than the Hubble horizon, such perturbations might grow sufficiently for strong gravity effects to become significant at horizon scales. In this case, the FLRW metric may no longer be an accurate description of the spacetime: such configurations could either lead to a spacetime dominated by radiation-like perturbations or, alternatively, strong backreaction on the geometry might lead to gravitational collapse and black hole formation. 

Depending on the abundance and amplitude of super-horizon perturbations, the formation of such primordial black holes (PBHs) may play an important role in cosmology, as it could lead to a period of early matter domination after kination with observational consequences. The critical value of the overdensity associated with a curvature perturbation at the time it enters the horizon, $\delta_c$, is essential to the calculation of such a PBH abundance (see e.g.~\cite{Heurtier_2023,Bhattacharya_2020}). The value of this critical overdensity was studied analytically using a {\it three-zone} solution (a spherically symmetric top-hat overdensity surrounded by a matching region, embedded in an FLRW universe) by~\cite{Harada_2013} and has the following form 
\begin{equation}\label{deltacritical}
    \delta_c=\frac{3(1+w)}{5+3 w}\sin^2\left(\frac{\pi\sqrt{w}}{1+3 w}\right)\,,
\end{equation}
which, for pure kination, corresponds to $w=1$ and $\delta_c = 0.375$. One of the primary goals of what follows is to test the robustness of this analytical result using numerical relativity.

\subsection{Initial Scalar Data}
Our simulation domain is a 3-dimensional box with a length $L$ and periodic boundary conditions. We adopt a ``pseudo-isotropic'' setup with a sinusoidal inhomogeneity of the scalar field and its velocity in all three spatial directions while keeping the kinetic background homogeneous:
\begin{equation}
    \phi_{0}=\bar\phi_0+\frac{\Delta\phi}{3}\sum^3_{i=1}\cos(\frac{2\pi N_{\phi,0} x_i}{L})\,, \label{eqn:profile}
\end{equation}
\begin{equation}
    \dot\phi_0=\sqrt{\frac{3}{4\pi}}\mpl H_0 + \frac{\Delta\dot\phi}{3}\sum^3_{i=1}\cos(\frac{2\pi N_{\dot\phi,0} x_i}{L}+\theta)
\end{equation}
where the subscript $0$ denotes a quantity evaluated at initial time, $x_i$ labels the spatial coordinates on the hyperslice evolving by the time coordinate $t$, $\bar\phi_0$ is the initial homogeneous scalar field, $\Delta\phi$ and $\Delta\dot\phi$ are the amplitudes of the sinusoidal scalar and kinetic inhomogeneities, $N_{\phi,0}$ and $N_{\dot\phi,0}$ are the mode numbers of the initial scalar field and kinetic inhomogeneity, $H_0$ is the initial Hubble parameter calculated in the absence of the inhomogeneity, i.e. in the pure kination limit, $\theta$ is the phase difference between the scalar and kinetic inhomogeneities. In this work, we have chosen to study the fate of a single self-perturbation mode of a scalar field. A thorough treatment of a full spectrum of modes is beyond the scope of this paper and is left for future investigation.

Since $V\ll\frac{1}{2}\dot\phi^2$ during a kination era, we consider the simple case where $V(\phi)=0$\footnote{Note that for this reason, we would not observe in our work the tracker solution described in~\cite{mosny2025selftrackingsolutionsasymptoticscalar} due to the absence of a potential.}. Therefore, the dynamics are insensitive to translations in field space, and we are allowed to choose the initial background field value to be $\bar\phi_0=0$. Since we impose periodic boundary conditions, the size of the simulation domain, $L$, is chosen to be the initial scalar field inhomogeneity wavelength, which is related to the initial Hubble length via
\begin{equation}\label{box}
    L\equiv (N_{\phi,0}H_0)^{-1}=(N_{\phi,0}\dot{\phi}_0)^{-1}\sqrt{\frac{3}{4\pi}}\mpl\,.
\end{equation}
With this setup, we define the gradient energy density as
\begin{equation}
    \rhog\equiv\frac{1}{2}\gamma^{ij}\partial_i \phi\partial_j\phi\,.
\end{equation}
We can compute the volume averages at the initial time to be 
\begin{equation}
    \langle \rhogo\rangle=\frac{\Delta\phi^2\pi^2}{3L^2}\,,
\end{equation}
and
\begin{equation}
    \langle\rhodko\rangle\equiv\langle\frac{1}{2}\langle\dot\phi_0^2\rangle-\frac{1}{2}\langle\dot\phi_0\rangle^2\rangle=\frac{1}{12}(\Delta\dot\phi)^2
\end{equation}
where $\langle \rangle$ denotes the volume average over the spatial hyperslice, defined as
\begin{equation}
    \langle  X\rangle\equiv\frac{\int_\Sigma d^3x\sqrt{\gamma}X}{\int_\Sigma d^3x\sqrt{\gamma}}\,.
\end{equation}
We define the initial overdensity parameter as
\begin{equation}\label{ratio}
    \delta_{0}\equiv\frac{\langle \rhogo\rangle+\langle\rhodko\rangle}{\rhobo}\,,
\end{equation}
which indicates the non-linearity of the initial inhomogeneity and allows us to set the amplitude of the inhomogeneities accordingly. In this work, we only consider the presence of one type of initial perturbation at a time and leave the study of initial perturbation mixing for future works.

\subsection{Initial Geometric Data}\label{geometry}
We foliate our spacetime following the ADM formulation with the metric
\begin{equation}
    ds^2=-\alpha^2dt^2+\gamma_{ij}(dx^i+\beta^idt)(dx^j+\beta^jdt)\,,
\end{equation}
where the quantities $\alpha$ and $\beta^i$ are the lapse and shift functions that are evolving as gauge choices encoding the coordinate freedom of general relativity, while $\gamma_{ij}$ is the 3-dimensional spatial metric of the spatial hyperslice. The spatial metric and the gauge quantities are evolving along with the extrinsic curvature $K_{ij} = \partial_t\gamma_{ij} + D_{(i}\beta_{j)}$, with $D_i$ the 3-dimensional covariant derivative with respect $\gamma_{ij}$, which acts as the canonical momentum of the metric in the Hamiltonian formulation of General Relativity. 

The extrinsic curvature is then further decomposed into its trace and traceless part:
\begin{equation}
    K_{ij} = A_{ij} + \frac{1}{3}\gamma_{ij}K\,,
\end{equation}
where the trace part $K$ is associated with the local expansion or contractions of spacetime. The traceless term $A_{ij}$ can be further decomposed into traverse and longitudinal components where the former is the usual ``gravitational wave'' degrees of freedom in the linear limit. The spatial metric can be also be decomposed as $\gamma_{ij} = \chi^{-1}\tilde{\gamma}_{ij}$ with $\chi$ being the conformal factor such that $det(\tilde{\gamma})=1$.

We impose periodic boundary conditions, which means that we set the initial size of the simulation domain to a characteristic scale and the universe consists of infinitely many such boxes. In the limit of FLRW spacetimes, the Hubble length grows as $H^{-1}\sim a^{3}$ during kination; while the wavelength of a single mode stretches proportionally to $a$. Therefore, while the mode is stationary in the simulation domain, its characteristic wavelength is decreasing relative to the Hubble length due to the size of the domain shrinking compared to the physical Hubble volume.

Given an initial scalar field configuration, the initial data for the hyperslice is solved with the CTTK method~\cite{Aurrekoetxea:2022mpw,Aurrekoetxea:2025kmm},  where we assume an initially conformally flat with spatial metric $\tilde{\gamma}_{ij} = \delta_{ij}$. For the initial hyperslice we assume $\alpha_0=1$, $\beta_0^i=B_0^i=0$, and fix the conformal factor $\chi = 1$ everywhere. This leads to an algebraic equation for the Hamiltonian constraint and a Poisson-like equation for the momentum constraint, which are solved to obtain the initial profiles of the local expansion parameter $K$ and trace-free extrinsic curvature $A_{ij}$.

Additionally, the use of periodic boundary conditions enforces integrability conditions on the constraint equations. Specifically, the integral of the constraints over the simulation domain must vanish. Setting $\chi =1$ implies that $\int d^3x \Pi\partial_i\phi =0$, at the initial slice.
\subsection{Evolution and Diagnostics}
We evolve the system of the geometric and matter quantities with the CCZ4~\cite{Dumbser_2018} evolution scheme, see details in the~\cite{Andrade_2021, Radia_2022}.

In the FLRW limit, the Hamiltonian constraint equation is identified with the Friedmann equation, where  Hubble parameter is the trace of the extrinsic curvature $H=-\frac{1}{3}K$. Throughout this work except for explicit mentioning, we are using a coarse-grained measure of the expansion, the volume-averaged conformal factor:
\begin{equation}
    a \approx \langle \chi\rangle^{-1/2}\,,
\end{equation}
which naturally translates to the scale factor in FLRW limit.

For the evolution of the gauge quantities, we choose a modified version of moving puncture gauge~\cite{Campanelli:2005dd, Baker:2005vv} where the lapse is evolving as:
\begin{equation}\label{lapse}
    \partial_t\alpha=-2\alpha(K-\langle  K\rangle-2\Theta)+\beta^i\partial_i\alpha,
\end{equation}
This gauge choice maintains the lapse function roughly at its initial value for relatively homogeneous expanding regions while approaching the $1+$log form in collapsing regions with $\langle K\rangle<0$. It thus better regularises the behaviour of the lapse function in a cosmological spacetime. We adopt the standard Gamma-driver condition
\begin{equation}
    \partial_t\beta^i=\frac{3}{4}B^i,
\end{equation}
\begin{equation}
    \partial_tB^i=\partial_t\Tilde{\Gamma}^i-B^i.
\end{equation}

In numerical relativity, we use the gauge parameters to define the conjugate momentum of the scalar field as
\begin{equation}
    \Pi\equiv\frac{1}{\alpha}(\dot{\phi}-\beta^i\partial_i\phi).
\end{equation}
This allows us to define the volume-averaged perturbation energy density as
\begin{equation}\label{perturbation}
    \langle \rhop\rangle\equiv\langle \rhog\rangle+\langle \rhodk\rangle,
\end{equation}
with
\begin{equation}
    \langle\rhodk\rangle\equiv\langle\rhok\rangle-\rhob
\end{equation}
\begin{equation}
    \langle \rhok\rangle\equiv\frac{1}{2}\langle \Pi^2\rangle,
\end{equation}
\begin{equation}
    \rhob\equiv\frac{1}{2}\langle \Pi\rangle^2.
\end{equation}
The growth of perturbation with respect to the kination background, explained above could eventually lead to a significant local overdensity. Such local overdensity could reach above a critical point and lead to the formation of a black hole. The formation and presence of a black hole is tracked by calculating if an apparent horizon has formed in the hyperslice~\cite{Thornburg_2003}. 

We calculate the masses of the black holes by measuring the minimum area of the apparent horizon. As there is no initial angular momentum in the system, we can approximate the black hole mass with the Schwarzschild radius, $M_{\text{BH}}^2\approx r^2_{\text{BH}}/4\approx A_{\text{AH}}/16\pi$. In geometric units, $[M]\sim[L]\sim[T]$. We will denote the Hubble scale at the start of the kination period as $H_{\text{kin},0}$. We define our geometric mass according to~\cite{Aurrekoetxea_2025}
\begin{equation}
    H_{\text{kin}, 0}=H_0\left(\frac{Mg}{\mpl}\right)^{n},
\end{equation}
with $H_0=1/(N_{\phi,0}L)$ defined in Eq.~\eqref{box}, and $n=-1$ denotes the dimensionality of $H$ in geometric units. The masses of the black holes in Planck units are then converted in a similar fashion with 
\begin{equation}
    M_{\text{BH,pl}}=M_{\text{BH,g}}\left(\frac{M_g}{\mpl}\right).
\end{equation}

\section{Numerical Results}\label{sect:results}
For our simulations, we use the multi-purpose numerical relativity code \grchombo~\cite{Clough_2015,Andrade_2021}. We tune the initial inhomogeneity amplitude and the field momentum based on the $\delta_{c,0}$ defined in Eq.~\eqref{ratio}. Our goal is to investigate the fate of the spacetimes when large scalar inhomogeneities emerge in a kination  background. We present our results based on the initial characteristic wavelengths of the inhomogeneities. Perturbations with $\lambda_0< H_0^{-1}$ constitute initially sub-horizon inhomogeneities and conversely, $\lambda_0> H_0^{-1}$, super-horizon inhomogeneities. 

\subsection{Sub-Horizon and Near-Horizon Scale Perturbations}
In this section, we investigate the effects of both large and small amplitude inhomogeneities in the sub-horizon regime, where $\lambda < H_0^{-1}$. In the perturbative limit, sub-horizon modes evolve as $\rhop \sim a^{-4}$, indicating a behaviour that resembles radiation. According to~\cite{eröncel2025universalbounddurationkination}, this behaviour is robust regardless of whether the inhomogeneities are in the linear or non-linear regime in the presence of a fixed background. As stated above, one normally expects the presence of a subdominant potential term, which may nevertheless remain subdominant in the post-inflationary evolution compared to the kinetic energy density and can therefore be safely neglected.

\begin{figure}[h]
\centering
\includegraphics[width=0.7\linewidth]{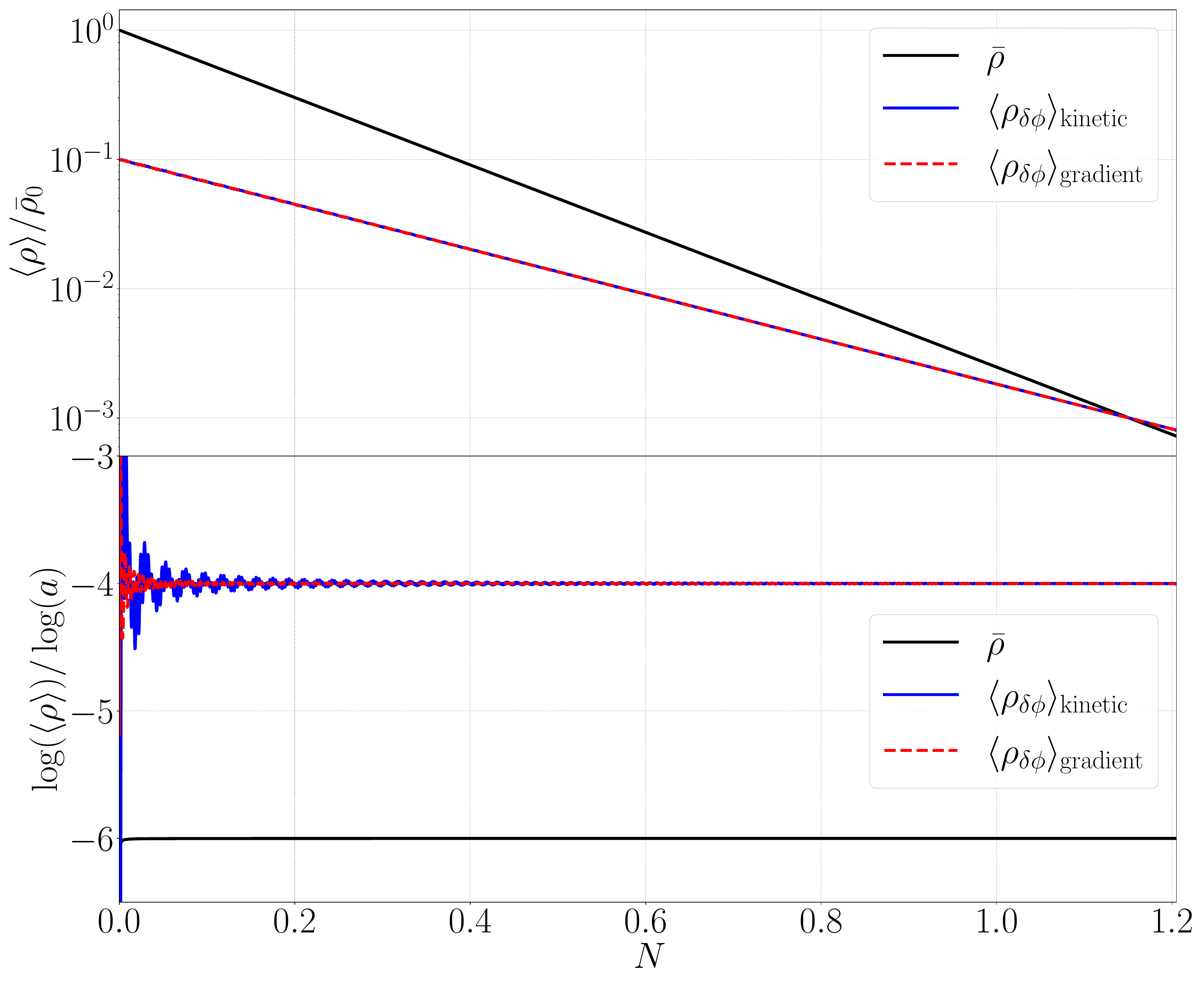}
\caption{Simulation results with initial gradient and kinetic perturbation in red and blue, respectively, for $\lambda_0=0.01H_0^{-1}$, $\delta_{0}=0.1$ -- at $N\sim 1.2$, the ``perturbation'' energy $\rho_{\delta \phi}$ dominates over the ``background'' energy density, but the two ``components'' still evolve as $a^{-4}$ and $a^{-6}$ respectively. The source of the initial perturbation energy density is indistinguishable to the system as long as they are of sub-horizon.  \textbf{Upper panel}: the evolution of volume-averaged energy densities of the background and perturbation. \textbf{Lower panel}: background and perturbation energy densities as a power law of the volume-averaged scale factor $a\approx\langle \chi\rangle^{-1/2}$.}\label{subhorizon}
\end{figure}

We simulate a scalar field with an initially homogeneous kinetic energy profile, superimposed with an inhomogeneity characterised by an initial wavelength of $\lambda = 0.01H_0^{-1}$ as in Eq.~\eqref{eqn:profile}. The amplitude of this inhomogeneity is chosen such that the volume-averaged initial gradient energy density is one tenth of the background kinetic energy density, i.e. $\delta_{0}=0.1$. To focus on the localised dynamics, we capture only a single full wavelength of the perturbation within our simulation domain and apply periodic boundary condition instead of the entire Hubble volume. We present our results in Figure~\ref{subhorizon}.

We found that a sub-horizon mode indeed behaves as $\langle \rhop\rangle\sim a^{-4}$, and would become $\mathcal{O}(1)$ with the kination background, and dominates as an effective radiation-like background. The continuity in the energy density profiles in Figure~\ref{subhorizon} indicates that the transition from kination domination to radiation domination is smooth. 

As~\cite{de_Jong_2022,deJong:2023gsx} has shown that critical collapse to a primordial black hole is possible with a supercritical sub-horizon initial overdensity during matter domination, we investigate such processes for kination. We discovered that for inhomogeneities with characteristic wavelengths of $\lambda\sim\mathcal{O}(10^{-1})H_0^{-1}$ the critical overdensity is of $\mathcal O(10^2\sim10^3)$. In such conditions, the universe's evolution is dominated by a highly localised gradient energy density of a massless scalar field. Therefore, gravitational collapse via accretion is highly suppressed. We thus conclude that, even if sub-horizon perturbations manage to grow during a kination era to become $\mathcal O(1)$, it is still very unlikely for such perturbations to undergo gravitational collapse.

As we probe sub-horizon inhomogeneities with wavelengths closer to the horizon size, the critical value $\delta_{0,\text{crit}}$ for black hole formation is reduced. For $\lambda_0 = 0.5H_0^{-1}$, we find $\delta_{0,\text{crit}} \approx 20$, which still implies that gradients must initially be very dominant. Once black holes form in our simulation, the universe becomes matter-dominated, with black holes forming on a lattice due to the symmetries enforced by the periodic boundary conditions.

In the near-Hubble limit, where $\lambda_0 = H_0^{-1}$, we find that black hole formation can occur more easily, with a critical value $\delta_{0,\text{crit}} \approx 2$. We note that while the wavelength is nominally of Hubble size, the significant amount of energy density from the gradient of the scalar field effectively shrinks the actual Hubble sphere, rendering the inhomogeneity effectively super-horizon. In the limit of such large inhomogeneities at this length scale, it is not meaningful to associate the energy density evolution with the scale factor, nor is the scale factor itself well-defined due to the lack of homogeneity. In cases where black holes do not form, we observe that the energy density eventually, on average, evolves as in a radiation-dominated universe. Additionally, for near-Hubble wavelengths, with $\delta_0<1$, the perturbations grow relative to the kination background and eventually dominate the evolution, again behaving like a radiation-dominated universe in the same manner as in the deep sub-horizon case.

To conclude this section: in the limit where a kination background with small perturbations is present, the perturbation theory result remains valid. On the other hand, when the initial $\delta_0 \gtrsim 1$, inhomogeneities dominate the evolution, but the gravitational backreaction is not strong enough to induce large-scale collapse. Finally, in the limit $\delta_0 \gg 1$, gravitational collapse can occur as sufficiently large inhomogeneities will always dominate local dynamics.

\subsection{Super-Horizon Gradient Perturbations}
\begin{figure}[!htbp]
    \centering
    \includegraphics[width=0.7\linewidth]{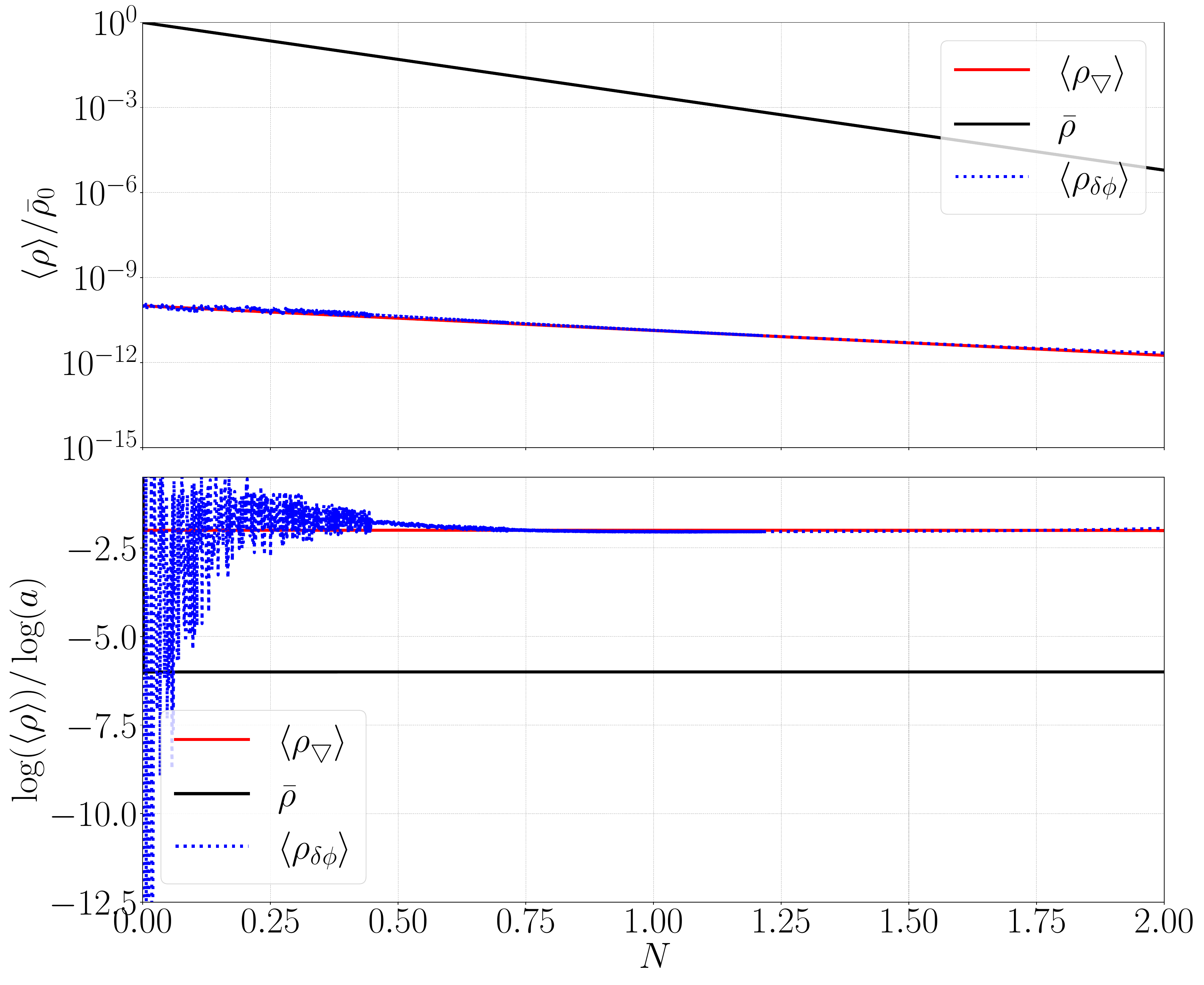}
    \caption{Simulation results for $\lambda_0=1\times10^3H_0^{-1}$, $\delta_{0}=1\times10^{-10}$. The perturbation energy density is closely tracked by the gradient energy density and evolve as $\langle \rhop\rangle\sim\langle \rhog\rangle\sim a^{-2}$ while the kination background remains at $\rhob\sim a^{-6}$. \textbf{Upper panel}: the evolution of volume-averaged energy densities of the background, the gradient, and perturbations, \textbf{Lower panel}: background, gradient, and perturbation energy densities as a power law of the volume-averaged scale factor $a\approx\langle \chi\rangle^{-1/2}$. }\label{superhorizon}
\end{figure}

For inhomogeneities with characteristic wavelength $\lambda_0>H_0^{-1}$, linear perturbation theory indicates that the perturbation energy density of super-horizon modes is approximately given by the gradient term and scales as $\rhop\sim\rhog\sim a^{-2}$. Thus, the density contrast grows as $a^4$ compared to the kination background and the modes behave like scalar curvature perturbations. We begin by confirming this in the linear limit and simulate a scalar field with a homogeneous kinetic energy profile and an inhomogeneity characterised by an initial wavelength of $\lambda_0=1\times10^3H_0^{-1}$. The amplitude of this inhomogeneity is chosen by setting the initial density contrast to $\delta_0=1\times10^{-10}$. The results of the simulation are shown in Figure~\ref{superhorizon}. We find that the volume-averaged perturbation energy density $\langle\rhop\rangle$ very quickly tracks the gradient energy density once the system
relaxes from the initial conditions. This happens within approximately one \efold. We also find that the variance of the local expansion parameter $\langle \delta K\rangle/\langle  K\rangle\ll1$, indicating that the simulation domain remains highly homogeneous and the spacetime closely follows the FLRW metric.  Thus, the perturbation energy density then evolves as $\langle \rhop\rangle\sim\langle \rhog\rangle\sim a^{-2}$.

\begin{figure}[!htbp]
    \centering
    \includegraphics[width=0.7\linewidth]{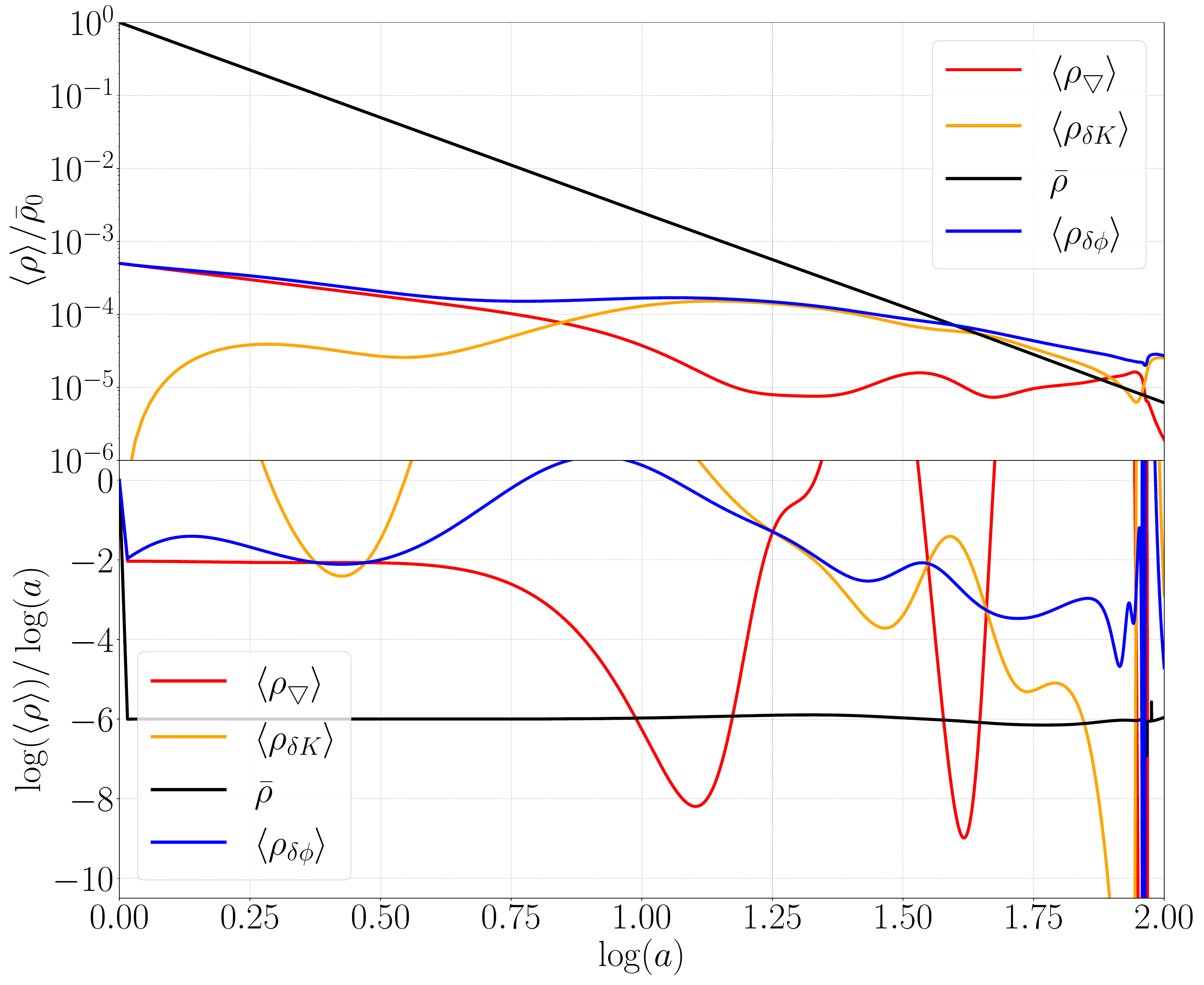}
    \caption{Simulation results for $\lambda_0=20H_0^{-1}$, $\delta_{0}=5\times10^{-4}$. The formation of black hole is indicated by the discontinuity around $\log(a)\sim 1.9$ at the end of the plot. The perturbation energy density deviates upwards away from the perturbative description when the induced kinetic perturbation becomes comparable with the gradient perturbation energy density. \textbf{Upper panel}: the evolution of volume-averaged energy densities of the background, gradient, kinetic, and perturbations, \textbf{Lower panel}: The background, gradient, kinetic, and perturbation energy densities as a power law of the $\log$ of the volume-averaged scale factor $a\approx\langle \chi\rangle^{-1/2}$. Simulation movie: \href{https://www.youtube.com/watch?v=0msEzENelt8}{https://www.youtube.com/watch?v=0msEzENelt8}}\label{nonlinear}
\end{figure}

Since the oscillation period of a super-horizon mode is proportional to its characteristic wavelength, $T \propto \lambda \gg H^{-1}$, the mode must oscillate and eventually re-enter the horizon before it can potentially enter a strong-gravity regime. To facilitate this within a feasible simulation time, we reduce the initial characteristic wavelength and increase the initial gradient energy density. As a representative example, we consider the case $\lambda_0 = 20H_0^{-1}$ and $\delta_0 = 5 \times 10^{-4}$, shown in Figure~\ref{nonlinear}.

We observe that $\langle \rhop\rangle\sim a^{-2}$ initially, the behaviours of both $\langle \rhop\rangle$ and $\langle \rhog\rangle$ quickly deviate from a constant power law once the oscillatory behaviour of the super-horizon mode and the induced kinetic perturbation energy density becomes significant. In particular, the behaviour for the total perturbation evolves approximately as $\langle \rhop\rangle\sim a^{-1}$, while the gradient term exhibits an oscillatory behaviour around $\langle \rhog\rangle\sim a^{-2}$. This indicates that the oscillatory behaviour and strong gravitational backreaction are significant.

Since during kination, the Hubble length redshifts as $H^{-1}\sim a^3$, while the wavelength of the mode redshifts as $\lambda\sim a$, we can approximate the horizon re-entry at
\begin{equation}
    a_{\text{re-entry}}\approx\sqrt{\lambda_0H_0}\,.
\end{equation}
This estimate assumes a consistent and well-defined notion of the scale factor throughout the simulation domain and evolution. To ensure this, we verify that most of the simulation domain remains well-described by an FLRW spacetime. Following the methods of~\cite{Giblin_2016, Aurrekoetxea_2025} we evaluate the variance, $\sigma^2$, of the local expansion parameter. In scenarios where black hole formation is expected, we additionally monitor the variance of the conformal factor and compute their kurtoses, $g_2$\footnote{We use the natural estimator definition of $$g_2=\frac{\frac{1}{n}\sum_i^n(x_i-\bar{x})^4}{\sigma^4}-3,$$ where $\sigma$ is the standard deviation. Under this definition, $g_2=-3$ indicates absolute uniformity, $g_2=0$ indicates a normal distribution, and $g_2>0$ indicates a sharp peak in distribution.}, to assess deviation from FLRW behaviour near the onset of gravitational collapse. In the simulation shown in Figure~\ref{nonlinear}, we observe that both $\sigma^2_K$ and $\sigma^2_\chi$ remain in the range $\mathcal O(10^{-3}\sim10^{-5})$ throughout the evolution. We also find that $g_{2,\chi}\sim g_{2,K}\sim-1$ prior to the black hole formation showing platykurtic(or near uniformity) distribution. These results confirm that the spacetime remains a good approximation to FLRW prior to the black hole formation.

In the simulation example shown in Figure~\ref{nonlinear}, we estimate the horizon re-entry to occur at $\log(a)_{\mathrm{re-entry}} \approx 1.5$ with $\langle \rhop\rangle\lesssim\rhob$ and black hole formation at $\log(a)_{\mathrm{BH}}\approx1.9$ as indicated by the discontinuity in the energy density profiles. The first detection of an apparent horizon and the estimated mode horizon re-entry are separated by approximately 0.5 \efolds. We observed that this behaviour is consistent across all super-horizon modes collapse we have tested. We note that while we can still attribute FLRW cosmological terms to the simulation, the simulation domain does deviate from a pure FLRW spacetime, making the precise moment of horizon re-entry difficult to define. 

\begin{figure}[!htbp]
    \centering
    \includegraphics[width=0.7\linewidth]{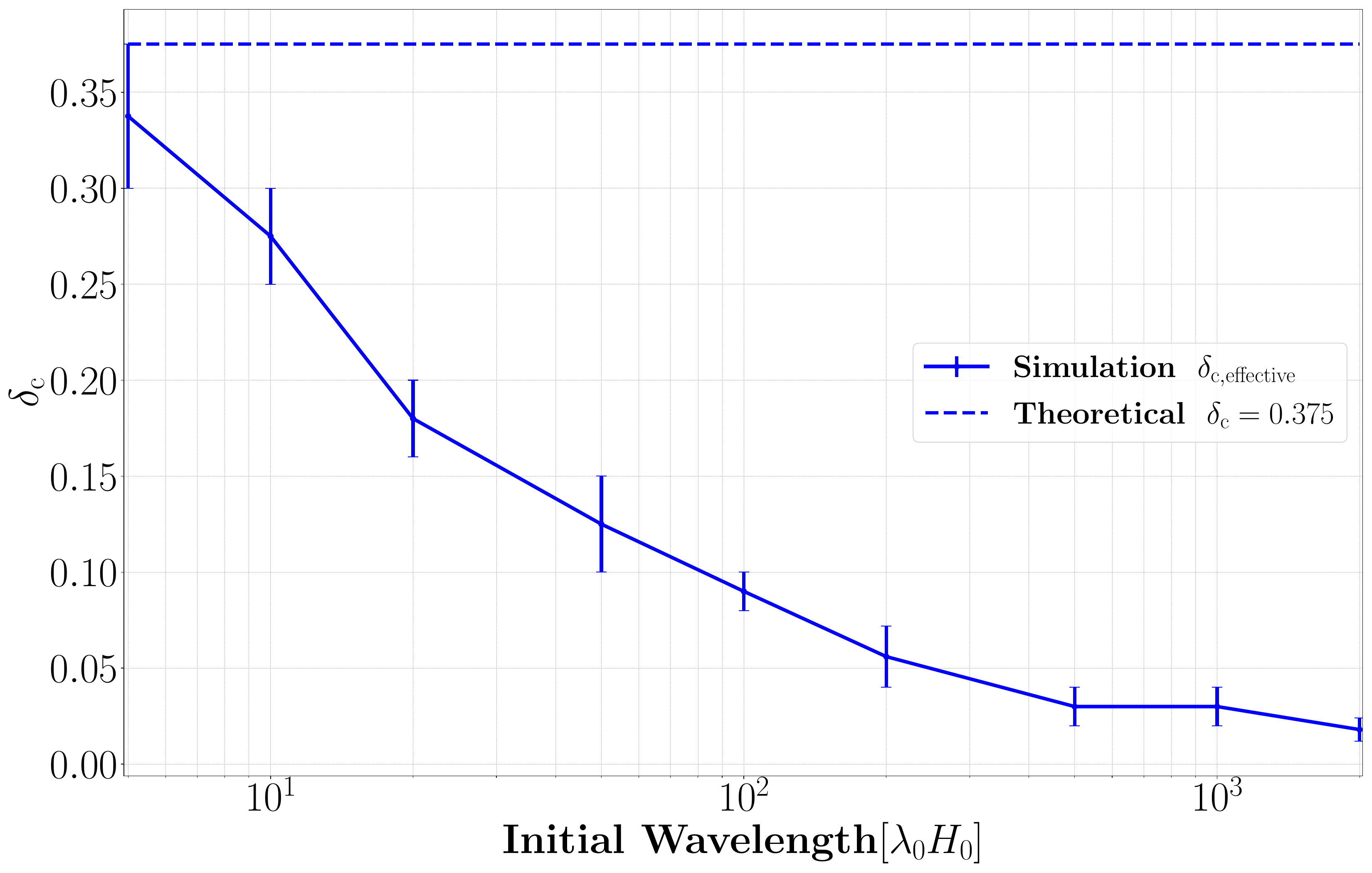}
    \caption{Extrapolated critical overdensity from numerical simulations for primordial black hole formation compared with the prediction from the Press-Schechter formalism of gravitational collapse~\cite{Bhattacharya_2020, Heurtier_2023}. The upper bound of the error bar corresponds the lowest amplitude tested with a black hole formation, whereas the lower bound of the error bar corresponds the highest amplitude tested without a black hole formation, and the solid blue line indicates the central value between the two bounds.}\label{fig:delta_c}
\end{figure}

We define $\delta_{c,\text{ effective}}$ as the effective critical overdensity, extrapolated from the simulation results under the assumptions that the kinetic energy density always dominates the universe's evolution prior to horizon re-entry, and that the perturbation grows relative to the kination background remains as $a^4$ 
\begin{equation}
    \delta_{c,\text{ effective}}\approx a^4\delta_{0,\text{crit}}=(\lambda_0H_0)^2\delta_{0,\text{crit}}
\end{equation}
Note that this quantity is defined for the sake of comparing our results to the perturbative case, but does not correspond to a quantity that is actually measurable numerically, unlike $\delta_{0,\text{crit}}$, which is the physical and measurable quantity in our approach. As mentioned above, we aim to investigate whether the critical overdensity is robust for super-horizon modes that exhibit non-linear behaviour before horizon re-entry. We simulated a range of super-horizon modes with initial wavelengths spanning from $5H_0^{-1}$ to $2000H_0^{-1}$ and an initial density contrast $\delta_0$ ranging from $\mathcal O(10^{-2}\sim10^{-10})$. We present the results in Figure~\ref{fig:delta_c}.

We emphasise again that the presented values do not represent the critical overdensity for primordial black hole formation at horizon re-entry. Instead, they serve as a comparison to the theoretical predictions if one expects the super-horizon modes to stay frozen as in the perturbative limit, until re-entry, i.e $\delta\sim a^4$. Under this assumption, these are the values that correspond to the critical initial conditions that lead to primordial black holes. We detail the cosmological implications in the next section.

\begin{figure}[!htbp]
    \centering
     \includegraphics[width=0.7\linewidth]{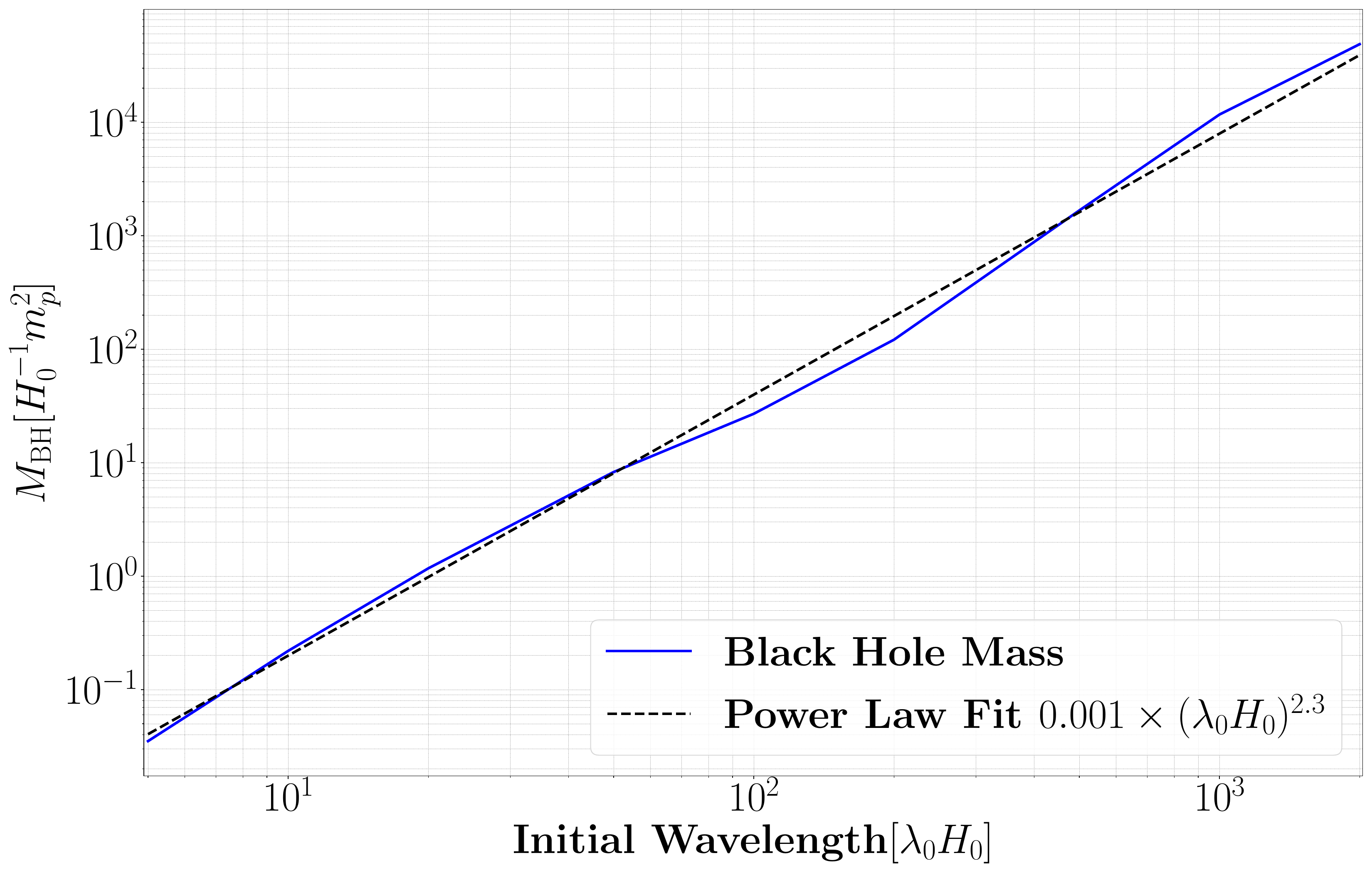}
    \caption{The mass of the black holes from simulations at the time of formation and a power law fit $M_{BH}\approx0.001\times(\lambda_0H_0)^{2.3}$}
    \label{MBH}
\end{figure}

In the simulation where black holes form, the lightest black holes at the point of formation are shown in Figure~\ref{MBH} with respect to the initial Hubble scale. To compare to the Hubble scale at the point of formation, we note that while the volume-averaged perturbation energy density is of $\mathcal O(1)$ compared to the background, the inhomogeneities are concentrated in the proximity of the primordial black hole both before and after formation. Therefore, we are still able to extract the Hubble parameter from the homogeneous bulk region. This corresponds roughly to $\mathcal O(10^{-2}\sim10^{-3})H^{-1}\mpl^2$.

The maximum initial wavelength we have tested in this work is $2000H_0^{-1}$. This limitation arises primarily due to the following factors 
\begin{enumerate}
\item As discussed above, we estimate the point of horizon re-entry occurs at $a_{\text{re-entry}}\approx\sqrt{\lambda_0H_0}$ with black hole formation typically taking place about $\sim0.5$ \efolds{} later. In this work, the largest scale factor we were able to probe is $a\sim80$. To investigate larger super-horizon modes that would re-enter the horizon during a prolonged kination era lasting up to $N_{\text{kin}}\sim10$, the simulation would need to reach scale factors of $\mathcal O(10^{3}\sim10^{4})$. However, achieving such large scale factors is currently infeasible with the numerical setup employed in this study. Numerical stability constrains the maximum allowed time step, and reaching significantly larger values of $a$ leads to an accumulation of numerical errors and requires computational resources that grow exponentially with simulation duration. 
\item The gauge choice we have imposed in Eq.~\eqref{lapse} ensures that the presence of a small perturbation does not induce strong gravitational backreaction until close to horizon re-entry. Thus the simulation domain is effectively a uniformly expanding spacetime with a decaying expansion rate. In such domains, we observe the local expansion parameter is $K\lesssim0$, with regions fluctuating slightly above and below the average. While Eq.~\eqref{lapse} is a more cosmology-friendly gauge~\cite{Aurrekoetxea_2025}, the prolonged oscillation periods for longer super-horizon modes gradually drive the lapse value across the entire hyperslice downward. As detailed in~\ref{geometry}, the lapse function is initially set to 1 everywhere on the hyperslice. The combined effect of the dynamics and this gauge choice causes the lapse function to slowly drop to $\mathcal O(10^{-1})$ across the domain, thereby increasing both the simulation time and computational cost.
\end{enumerate}

\subsection{Super-Horizon Kinetic Perturbations}
For completeness, we also investigated the effects of a pure kinetic super-horizon perturbation in this regime. For such a mode, perturbation theory suggests $\rhodk\sim a^{-6}$ (see Appendix. \ref{perturbationtheory}). However, as we have seen in the previous section that the super-horizon gradient perturbation redshifts as $a^{-2}$, the induced super-horizon gradient mode would shortly become the dominant contribution of the scalar perturbation. We present the result from the simulation with the same $\lambda_0$ and $\delta_0$ as in Figure \ref{nonlinear}. We observe that initially, the kinetic perturbation drops drastically to induce gradient energy until they are of the same order. The total perturbation energy density then behaves roughly as $a^{-2}$ before the estimated re-entry point. However, since the gradient energy density is induced at a later time than in the simulation of Figure \ref{nonlinear} and at a lower order, it does not have enough time to grow to the same non-linear regime as in the previous simulation, and thus do not undergo gravitational collapse. After the estimated re-entry at around $\log(a)\sim1.5$, the total scalar perturbation energy density redshifts as $a^{-4}$, which again confirms the result for the sub-horizon case. 
\begin{figure}[!htbp]
    \centering
    \includegraphics[width=0.7\linewidth]{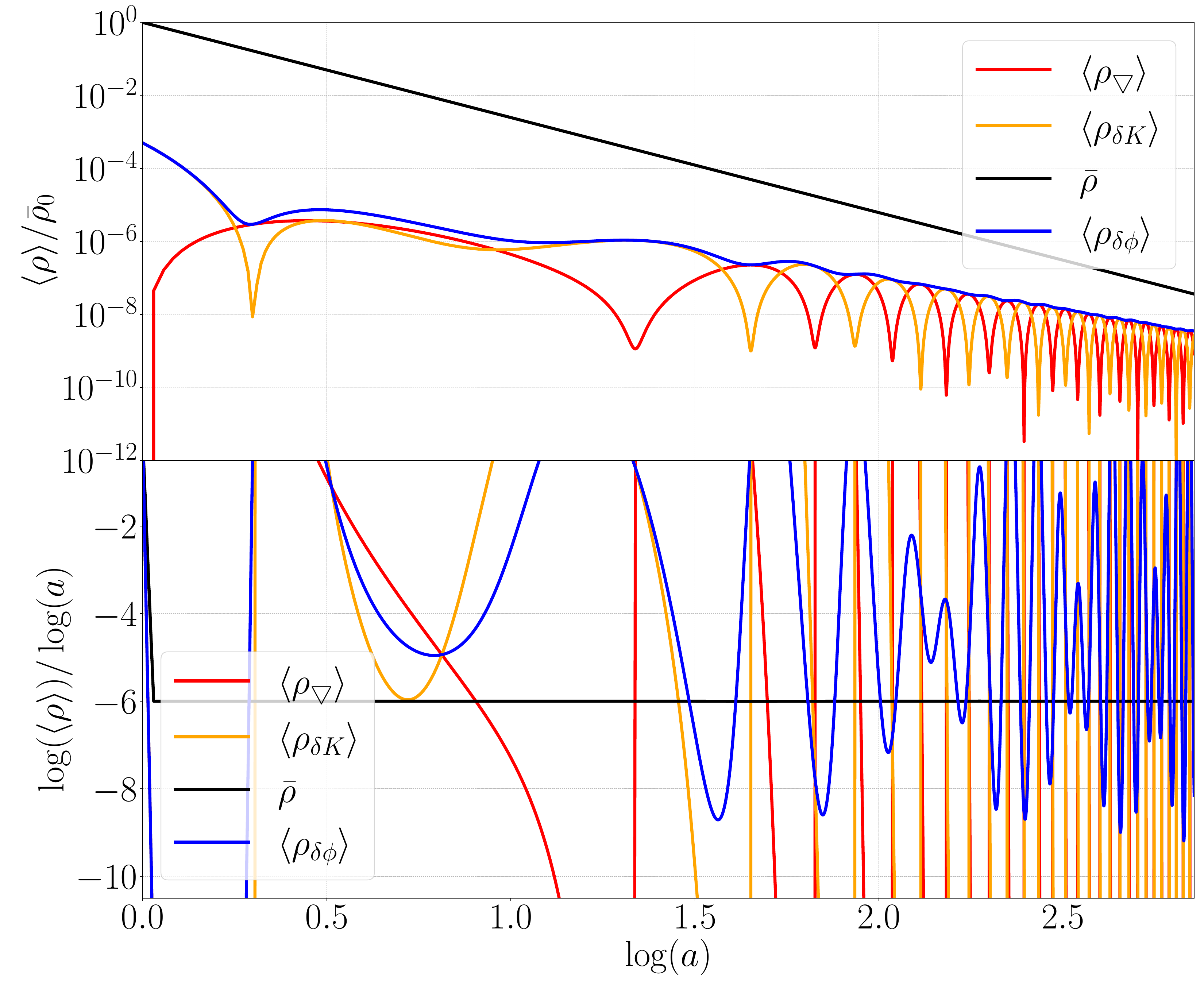}
    \caption{Simulation results for super-horizon kinetic perturbation with initial conditions of $\lambda_0=20H_0^{-1}$, $\delta_{0}=5\times10^{-4}$. \textbf{Upper panel}: the evolution of volume-averaged energy densities of the background, gradient, kinetic, and perturbations, \textbf{Lower panel}: The background, gradient, kinetic, and perturbation energy densities as a power law of the $\log$ of the volume-averaged scale factor $a\approx\langle \chi\rangle^{-1/2}$. }
    \label{superhorizonkinetic}
\end{figure}
\section{Implications for Cosmology}\label{sect:implications}
Scalar inflation models featuring a runaway direction in field space---and thus naturally leading to a kination era---generically face a challenge: how to reheat a universe dominated by the kinetic energy of a rolling massless scalar field? The most common answer to that question so far was to invoke the gravitational production of SM particles, generating a finite amount of radiation at the end of inflation that eventually overtakes the kination field (see, for instance,~\cite{Dimopoulos:2017zvq} and references therein). However, such a gravitational production is typically very inefficient, requiring a prolonged kination phase to be successful\footnote{This result is obtained assuming that the SM of particle physics contains $g_\star = 106.75$ relativistic degrees of freedom at the inflation energy scale, and taking the efficiency factor $q=1$ in Eq. (26) of~\cite{Dimopoulos:2017zvq}.}
\begin{equation}\label{eq:Nkin}
N_{\rm kin} \gtrsim  16.46 + \log \left(\frac{10^{14}\mathrm{GeV}}{H_{\rm end}}\right)\,,
\end{equation}
where $H_{\rm end}$ denotes the Hubble scale at the end of inflation. From this equation, and given existing upper bounds on the energy scale during inflation~\cite{Planck:2018vyg}, one finds that reheating the universe via gravitational reheating alone after kination requires $N_{\rm kin}\gg 10$. However, such a long period of kination is precisely what was shown recently to be very difficult--if not impossible to sustain--as scalar perturbations appear to dominate over the kination background after about 10 \efolds{}~\cite{gouttenoire2022kinationcosmologyscalarfields,mosny2025selftrackingsolutionsasymptoticscalar}. This suggests that an alternative mechanism is necessary to reheat the universe following kination.

There exist at least two other plausible mechanisms to reheat the universe after the kination phase ends: 
\begin{enumerate}
    \item The kination field will eventually develop a mass term, allowing it to decay and transfer its energy density to the radiation sector--as exemplified in~\cite{apers2024stringtheoryhalfuniverse}--or alternatively, a phase transition might abruptly end the kination period, or something else is produced during kination that succeeds in dominating the energy density of the universe above the Big-Bang Nucleosynthesis (BBN) temperature. 
    
    \item Primordial Black Holes (PBHs) formed during the kination era could serve this purpose in the second scenario: if produced during kination, they can rapidly dominate the energy density before evaporating and reheating the universe. Indeed, since they constitute a cold matter component of the universe prior to evaporation, PBHs would eventually dominate over both the kination background and the radiation contributions from the scalar perturbations. Whether such PBHs could reheat the universe thus depends on their initial abundance and lifetime, such that they evaporate only after they have begun dominating. 
\end{enumerate}
Let us henceforth assume that the primordial power spectrum of scalar curvature perturbations $\Delta^2_{\mathcal R}(k)$ features--beyond the nearly scale-invariant contribution $\Delta^2_{\mathcal R,\mathrm{CMB}}$ responsible for the CMB modes---a quasi-monochromatic enhancement at a specific wave-number $k_{\rm PBH}$, such that
\begin{equation}
    \Delta^2_{\mathcal R}(k) = \Delta^2_{\mathcal R,\mathrm{CMB}}(k) + \Delta^2_{\mathcal R, \mathrm{peak}}\delta(k-k_{\rm PBH})\,. 
\end{equation}
Typically, the power spectrum responsible for the perturbations observed in the CMB has an amplitude too small to produce any primordial black holes (PBHs), resulting in fewer than one PBH per Hubble patch. However, in the presence of features along the inflation trajectory that lead to brief phases of ultra-slow-roll such as is the case in~\cite{Heurtier_2023} within the context of kination-- the peak amplitude $\Delta^2_{\mathcal R, \mathrm{peak}}$ may become large enough to generate a significant fraction of PBHs. This energy fraction of the universe's energy density at the time of PBH formation, commonly denoted as $\beta$, is extremely sensitive to the critical overdensity $\delta_c$ above which perturbations collapse into black holes at horizon re-entry (see~\cite{Byrnes:2021jka}) and scale as
\begin{equation}
    \beta\sim \frac{\Delta_{\mathcal R, \mathrm{peak}}}{\sqrt{2\pi}\delta_c}\exp(-\frac{\delta_c^2}{2\Delta^2_{\mathcal R, \mathrm{peak}}})\,.
\end{equation}
Demanding that the PBHs form when mode $k_{\rm PBH}$ re-enters the horizon, we find that this fractions needs to satisfy
\begin{equation}\label{eq:conditionPBH}
    \beta\gtrsim \left(\frac{\Gamma_{\rm ev}}{H_{\rm end}}\right)^{1/2}e^{-\frac{3}{2}N_{\rm kin}+3(N_{\rm form}-N_{\rm end})}\,,
\end{equation}
where we chose $N_{\rm kin}\approx 10$ the total number of \efolds{} of kination until radiation-like scalar perturbations dominate over the kination background, $N_{\rm form}-N_{\rm end}$ denotes the number of \efolds{} between the end of inflation and PBH formation, and $\Gamma_{\rm ev}\approx 10^{-2}\mpl^4/M^3$ is the evaporation rate of PBHs with mass $M$.

\begin{figure}[!htbp]
    \centering
    \includegraphics[width=0.7\linewidth]{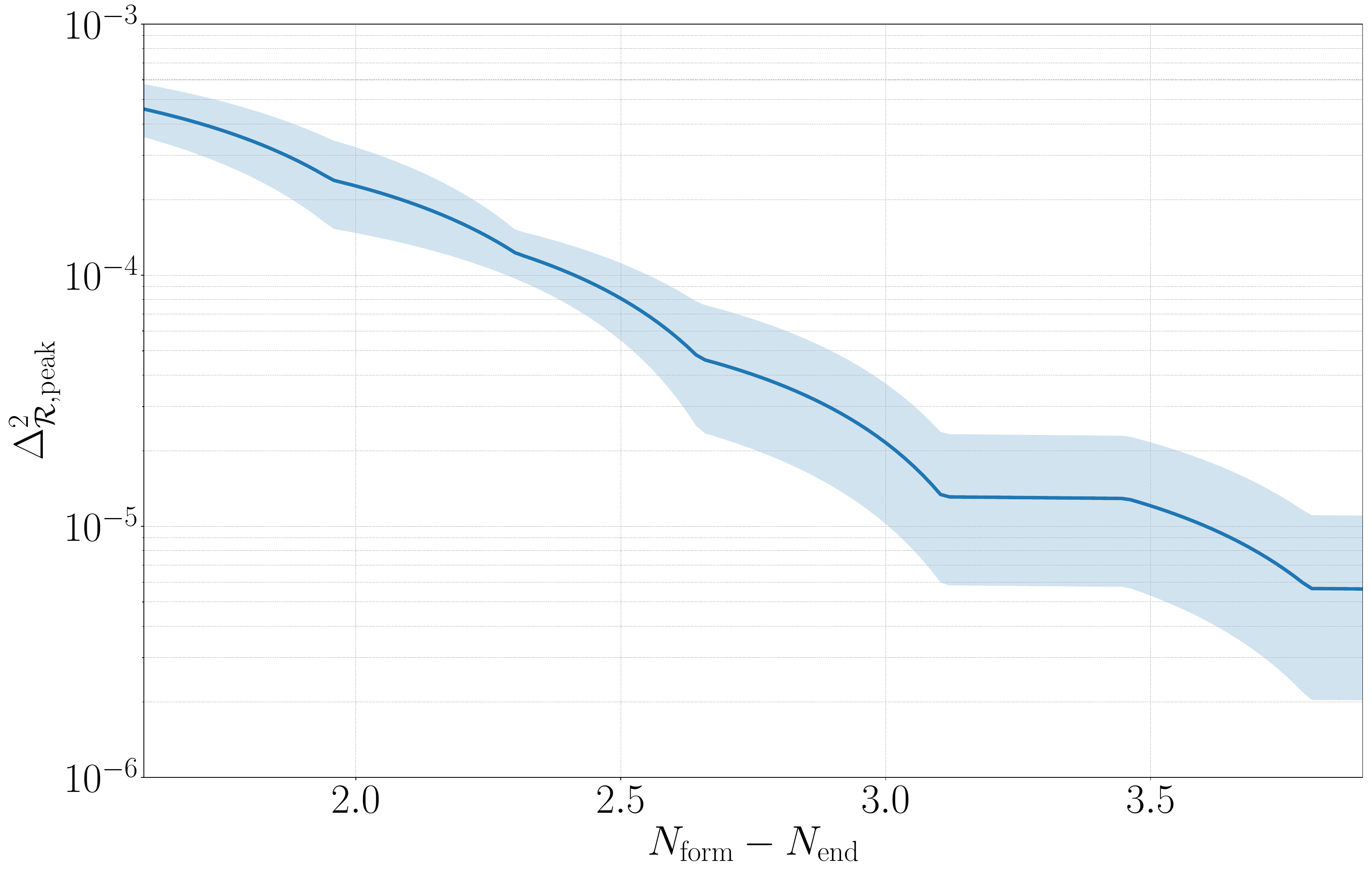}
    \caption{\label{fig:PS}Minimum peak value of the power spectrum of scalar curvature perturbation necessary to produce enough PBHs, $N_{\rm form}-N_{\rm end}$ \efolds{} after the end of inflation. The values of $\delta_c$ used to derive the blue central line and the shaded bands around it correspond to the central values and error bars depicted in Figure~\ref{fig:delta_c}.}
\end{figure}

In Figure~\ref{fig:PS}, we translate the results obtained in Section~\ref{sect:results} for $\delta_c$ into lower bounds on $\Delta^2_{\mathcal R, \mathrm{peak}}$ that satisfy the condition of Eq.~\eqref{eq:conditionPBH} for an arbitrary value of $H_{\rm end}=10^{13}\mathrm{GeV}$. Nevertheless, due to the logarithmic dependency of $\Delta^2_{\mathcal R, \mathrm{peak}}$ on $\beta$, we note that the results depend only very mildly on the actual value of $H_{\rm end}$. From the figure, it is clear that the very small values of $\delta_c$ we obtained for large perturbations (which form at larger $N_{\rm form}$) reduce the required amplitude of the power spectrum by more than an order of magnitude compared to previous claims in the literature that use the analytical results of~\cite{Harada_2013}.

This result is particularly important, as it is generally very challenging--from a model-building perspective--to enhance the power spectrum of comoving scalar curvature perturbations all the way up to $\Delta^2_{\mathcal R, \mathrm{peak}}\sim \mathcal O(10^{-2}\sim10^{-3})$ without requiring significant fine-tuning of the inflaton potential parameters~\cite{Cole:2023wyx}. Moreover, such enhancements often risk conflicting with the CMB predictions, especially regarding the spectral index $n_s$ and the tensor-to-scalar ratio $r$. However, according to our results, the required power spectrum amplitude during a kination era for PBHs to play a significant cosmological role only needs to reach $\sim\mathcal O(10^{-5})$. Achieving this level of enhancement within an inflationary scenario featuring a transient ultra-slow-roll phase does not suffer from the same fine-tuning issues, and is therefore more plausible from an ultraviolet complete theoretical framework.

\section{Summary and outlook}\label{sect: summary}
We have investigated the evolution of scalar inhomogeneities during a kination epoch beyond the linear regime using numerical relativity. Starting with small perturbations, our results in the sub-horizon limit agree with perturbative analysis: linear perturbations grow relative to the background and eventually dominate the universe, but do not collapse into black holes. We then examined initially large perturbations and found that they rapidly begin to behave like radiation unless the initial overdensity is large $\delta_0 \gg 1$, in which case gravitational forces are strong enough to induce immediate collapse into a black hole. However, such large sub-horizon perturbations cannot arise from the growth of initially small fluctuations after horizon re-entry, since once they dominate over the background, further growth is suppressed, thereby preventing collapse. We therefore conclude that only mechanisms capable of sourcing large sub-horizon inhomogeneities at early times could lead to gravitational collapse.

On the other hand, super-horizon inhomogeneities exhibit a much richer phenomenology with respect to strong-gravity effects. The very long oscillation timescales of these modes imply that they behave like curvature perturbations prior to horizon re-entry. We observed that even in the limit where $\delta_0 \ll 1$, such perturbations grow significantly relative to the background within just a few \efolds{} of kination. After horizon re-entry, their fate depends on their amplitude, they may either evolve into a radiation-like background--similar to typical sub-horizon modes--or collapse into black holes. Our key finding is that the critical overdensity extrapolated from our numerical evolution is much smaller than perturbative predictions, although it converges to the perturbative predictions when the initial wavelength approaches $H_0^{-1}$. For increasingly larger wavelengths, we find that the extrapolated critical overdensity decreases, reaching around $\delta_{c,\text{numerical}}\approx 0.02$ at $\lambda_0\approx2000H_0^{-1}$, as shown in Figure~\ref{fig:delta_c}.

This significant reduction of the extrapolated critical overdensity for the collapse of super-horizon modes into PBHs as compared to previously existing results in the literature is crucial. Indeed, the capacity of Gaussian perturbations---characterized by a primordial power spectrum $\Delta_\mathcal{R}^2
$--to produce PBHs from their horizon re-entry is exponentially sensitive to this critical value, $\beta \simeq \exp(-\delta_c^2/\Delta_\mathcal{R}^2)$. Reducing $\delta_c$ by over an order of magnitude thus considerably eases PBH formation. We estimated the typical value that would be required for $\Delta_\mathcal{R}^2$ for PBH formation during kination to lead to PBH domination and evaporation before BBN, and found that values of $\mathcal O(10^{-5})$ were sufficient--compared to around $\mathcal O(10^{-3})$ in the standard kination scenario.

It is also worth noting that in our numerical simulations, the requirement that the initial spatial hyperslice be conformally flat, combined with the imposition of periodic boundary conditions, enforces a high degree of symmetry in our physical setup. Intuitively, one might expect that in the case of black hole formation, the collapse of a single perturbation mode in three spatial directions might remove some energy from the system to create the spherical configuration that eventually collapses. Therefore, we cannot exclude the scenario in which the numerical value of $\delta_c$ in the case of direct spherical collapse might be even lower than what we have inferred in section~\ref{sect:results}. It is important that future simulations explore such spherically symmetric configurations so that the expression for $\delta_c$ can be verified more accurately; in this work, we did not pursue this direction due to the difficulty in finding initial conditions that both respect spherical symmetry and satisfy the integrability conditions imposed by periodic boundary conditions. On the opposite, additional degrees of non-sphericity might as well render the collapse more difficult to achieve~\cite{escrivà2024simulationsellipsoidalprimordialblack}, which also requires a specific study in the context of kination.

Finally, it is necessary to comment on the choice of gauge. In our numerical simulations, we employ a modified version of the moving puncture gauge, which includes a contribution from the volume-averaged extrinsic curvature scalar. As explained in Section~\ref{sect:results}, this particular gauge choice leads to the lapse function eventually dropping to very low values, effectively freezing the simulations for very large initial scalar wavelengths. A more suitable gauge could enable the evolution of the system for at least seven \efolds{} of kination, potentially allowing for a more accurate determination of $\delta_c$ in the limit of very large initial wavelengths. We leave such explorations to future work.

\section{Acknowledgements}
We acknowledge useful conversations with Martin Mosny, Joseph Conlon, Benjamin Stefanek, and Katy Clough. CC and PG thank Santiago Agüí Salcedo, Gonzalo Villa, Lucas Vicente Garcia-Consuegra, Josu Calvo Aurrekoetxea, Christian Byrnes, Robyn Munoz, Daniel G. Figueroa and Liina Chung-Jukko for their useful input. We would also like to thank the GRChombo team \href{http://www.grchombo.org}{(http://www.grchombo.org/)}. This work is supported by a Research Project Grant RPG-2021-423 from Leverhulme Trust.

This work was performed using the DiRAC@Durham
facility managed by the Institute for Computational
Cosmology on behalf of the STFC DiRAC HPC Facility (www.dirac.ac.uk) under DiRAC RAC15 Grant ACTP316. The
equipment was funded by BEIS capital funding via
STFC capital grants ST/P002293/1, ST/R002371/1 and
ST/S002502/1, Durham University and STFC operations
grant ST/R000832/1. This work also used the DiRAC
Data Intensive service at Leicester, operated by the University of Leicester IT Services, which forms part of the STFC DiRAC HPC Facility (www.dirac.ac.uk). The equipment was funded by BEIS capital funding via STFC capital grants ST/K000373/1 and ST/R002363/1 and STFC DiRAC Operations grant ST/R001014/1. DiRAC
is part of the National e-Infrastructure. The work of LH is supported by the STFC (grant No. ST/X000753/1).

\appendix
\section{Linear Perturbations During Kination}\label{perturbationtheory}

We briefly review the relevant solutions for scalar perturbations during kination on a fixed FLRW background. For a more detailed analysis, including the effects of metric perturbations, we refer the reader to~\cite{apers2024stringtheoryhalfuniverse,eröncel2025universalbounddurationkination,mosny2025selftrackingsolutionsasymptoticscalar}. The scalar field in the presence of small perturbations can be written as 
\begin{equation}
    \phi(\eta,\vec{x}) = \bar{\phi}(\eta) + \delta\phi(\eta,\vec{x}) = \bar{\phi}(\eta) + \frac{f(\eta,\vec{x})}{a}\,.
\end{equation}
The perturbation can be decomposed into Fourier modes as
\begin{equation}
   \delta\phi(\eta,\vec{x}) = \int \frac{d^3k}{(2\pi)^3}\delta\phi_{k}(\eta)e^{i\vec{k}\cdot\vec{x}}\,.
\end{equation}

The equations of motion for the background scalar field $\bar{\phi}$ and its perturbations in Fourier space, in the limit where the scalar potential vanishes ($V\to0$), are given by
\begin{align}\label{eqn:perteom}
     & {\bar{\phi}}'' + 2\mathcal{H}{\bar{\phi}}'= 0\,, \\
     & f_k'' + \left(k^2 - \frac{a''}{a}\right)f_k = 0\,,
\end{align}
where the perturbation equation is the Mukhanov--Sasaki equation. For a kination background, this equation becomes
\begin{equation}
     f_k'' + \left(k^2 + \frac{1}{4\eta^2}\right)f_k = 0\,.
\end{equation}

During kination, the scale factor evolves as
\begin{equation}
    a=a_0\left(\frac{\eta}{\eta_0}\right)^{1/2}\,,
\end{equation}
and the conformal Hubble parameter is
\begin{equation}
    \mathcal{H}=\frac{1}{2\eta}\,,
\end{equation}
while the background density evolves as
\begin{equation}
    \bar{\rho}\sim a^{-6}\,.
\end{equation}

In pure kination, assuming initial conditions $\bar{\phi}(\eta_0)=0$ and $\bar{\phi}'(\eta_0)=\Pi_0$, the background scalar field evolves as 
\begin{equation}
    \bar{\phi}(\eta) \sim \ln\left(\frac{\eta}{\eta_0}\right)\,.
\end{equation}

The general solution of the Mukhanov--Sasaki equation can be expressed in terms of Bessel functions $J_0$ and $Y_0$,
\begin{equation}
     f_k(\eta)=A\sqrt{\eta}\,J_0(k\eta)+B\sqrt{\eta}\,Y_0(k\eta)\,.
\end{equation}

We now investigate the asymptotic limits of this solution.

In the deep sub-horizon limit, $k\eta\gg1$, the asymptotic behavior of the Bessel functions leads to oscillatory solutions,
\begin{equation}
    \delta\phi_k = \frac{f_k}{a}\sim \frac{\cos(k\eta)}{a}\sim a^{-1}\,.
\end{equation}

We therefore conclude that the gradient energy density of the perturbations evolves as
\begin{equation}
    \langle\rho_{\nabla}\rangle\sim \left\langle\frac{\delta\phi_{k}^2}{a^2}\right\rangle\sim a^{-4}\,,
\end{equation}

while the kinetic energy density of the perturbations evolves as
\begin{align}
   \langle\rho_{\delta K}\rangle
   &= \left\langle\left(\frac{f_k'}{a^3}-\mathcal{H}\frac{f_k}{a^2}\right)\bar{\phi}'\right\rangle \notag\\
   &\quad + \left\langle\left(\frac{f_k'^2}{2a^4}
   + \mathcal{H}^2\frac{f_k^2}{2a^2}
   - \mathcal{H}\frac{f_k'f_k}{a^3}\right)\right\rangle
   \sim a^{-4}\,.
\end{align}

Therefore, the total perturbation energy density evolves as
\begin{equation}
    \rho_{\delta\phi} =\langle \rho_{\nabla} + \rho_{\delta K}\rangle\sim a^{-4}\,.
\end{equation}

The density contrast in the sub-horizon regime is therefore
\begin{equation}
    \delta_{\text{sub}} = \frac{\rho_{\delta\phi}}{\bar{\rho}}\sim a^2\,.
\end{equation}

In the super-horizon regime, $k\eta\ll1$, the solution becomes
\begin{equation}
    f_k = c_1\eta^{1/2} + c_2\eta^{1/2}\ln\eta
    \quad \Rightarrow \quad
    f_k = Aa + Ba\ln a\,.
\end{equation}

Consequently,
\begin{equation}
    \delta\phi_k = A + B\ln a\,.
\end{equation}

Note that, unlike the de Sitter case where $a''/a=2/\eta^2$ and the perturbations freeze completely outside the horizon, the scalar perturbations during kination evolve logarithmically. However, if the initial velocity satisfies $\delta\phi'(\eta_0)=0$, then $B=0$, eliminating the logarithmic contribution, and the mode remains frozen while outside the horizon.

In this case,
\begin{equation}
    \rho_{\delta\phi}\sim\langle\rho_{\nabla}\rangle\sim\frac{\delta\phi_{k,0}^2}{a^2}\sim a^{-2}\,.
\end{equation}

The conclusion is that deep sub-horizon modes grow relative to the kination background as
\begin{equation}
    \delta_{\text{sub}}\sim a^2\,,
\end{equation}
while super-horizon modes grow as
\begin{equation}
    \delta_{\text{super}}\sim a^4\,.
\end{equation}

If super-horizon perturbations in the kinetic sector are also included, the logarithmic term contributes nontrivially. However, we find that
\begin{equation}
    \langle\rho_{\delta K}\rangle\sim a^{-6}\,,
\end{equation}
which means that the kinetic perturbations decay at the same rate as the background. Therefore, the dominant contribution comes from the gradient energy density, which decays in the same way as scalar curvature perturbations.

\section{Convergence Tests}
\begin{figure}[!htbp]
    \centering
    \includegraphics[width=0.64\linewidth]{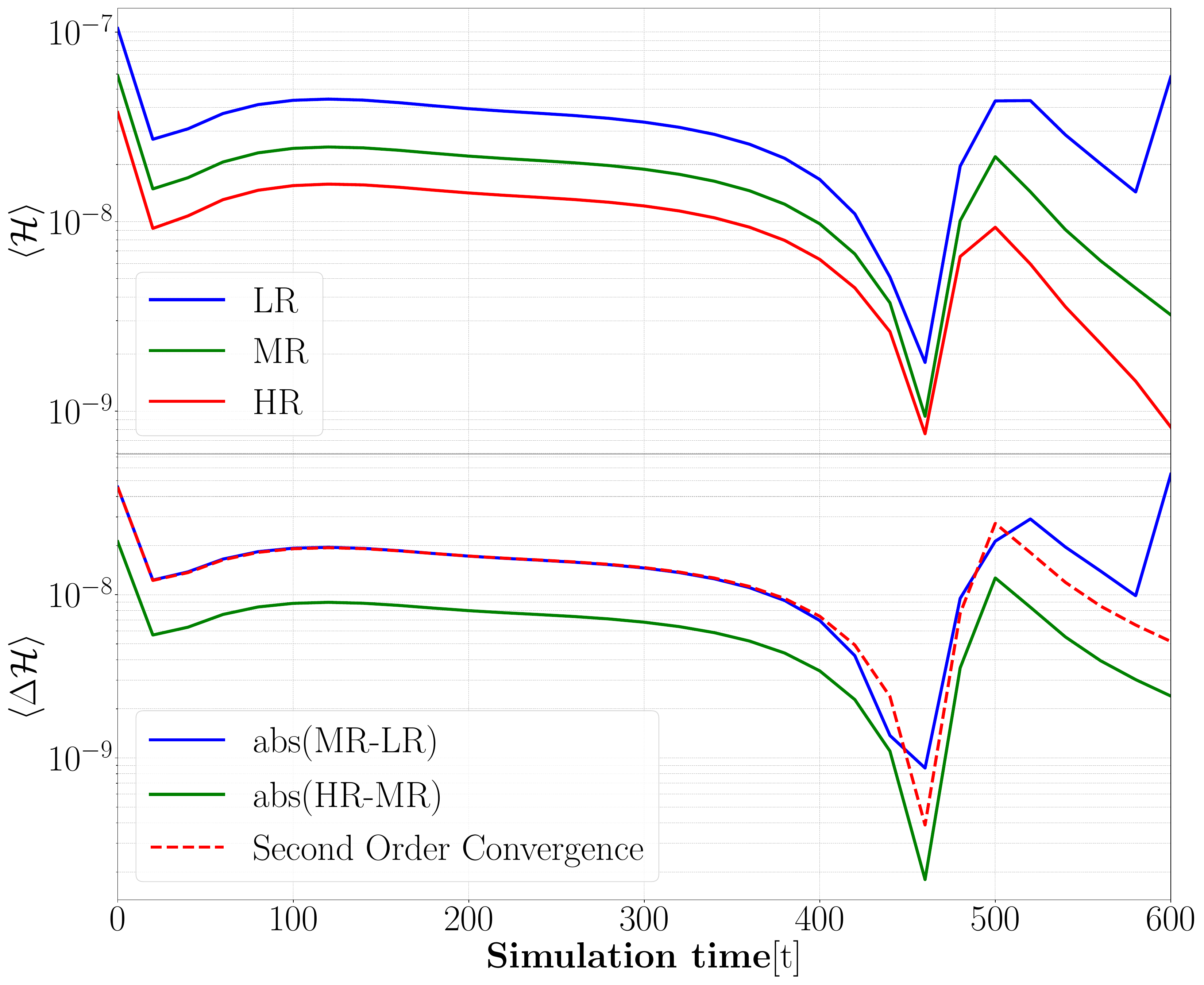}
    \caption{Hamiltonian convergence tests for volume-average Hamiltonian constraint violation with initial perturbation characteristic wavelength $\lambda_0=20H_0^{-1}$, and initial non-linearity parameter $\delta_0=5\times 10^{-4}$, \textbf{Upper Panel}: The volume-averaged Hamiltonian constraint with grid resolutions $N_{LR}=128$, $N_{MR}=144$, $N_{HR}=160$, \textbf{Lower Panel}: The difference in volume-averaged Hamiltonian constraint between high-middle and middle-low resolutions, and the second order convergence line from abs(HR-MR) to abs(MR-LR) according to Eq.~\eqref{converge}, showing good level of second order convergence.}
    \label{HamConvergence}  
\end{figure}

\begin{figure}[!htbp]
    \centering
    \includegraphics[width=0.64\linewidth]{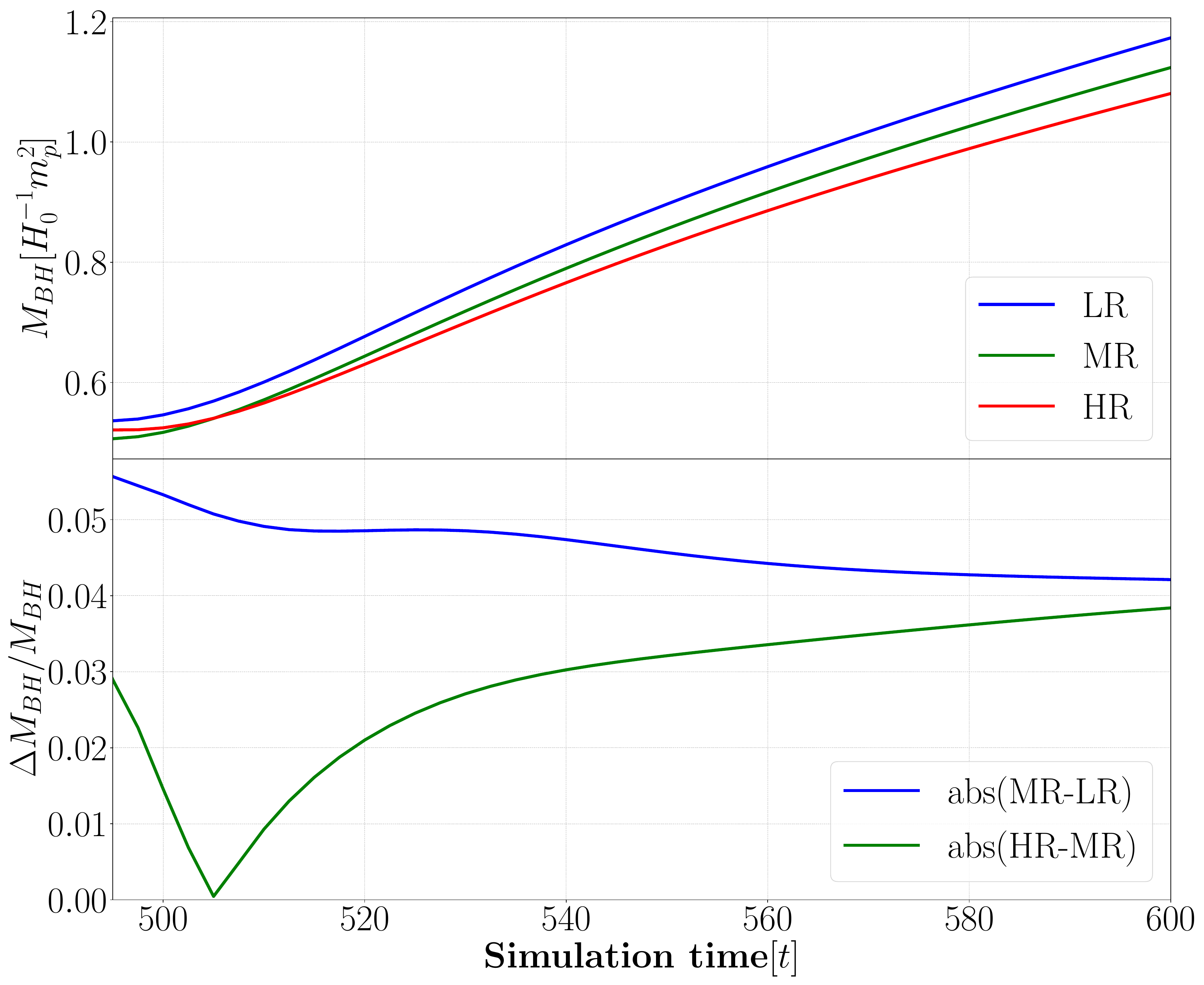}
    \caption{Convergence tests for black hole mass formed with initial perturbation characteristic wavelength $\lambda_0=20H_0^{-1}$, and initial density contrast $\delta_0=5\times 10^{-4}$, \textbf{Upper Panel}: The measured black hole mass with apparent horizon finder with grid resolutions $N_{LR}=128$, $N_{MR}=144$, $N_{HR}=160$, \textbf{Lower Panel}: Errors in black hole mass between high-middle and middle-low resolutions showing convergence to $<5\%$.}
    \label{MassConvergence}  
\end{figure}

We test the robustness of our numerical simulation by comparing the evolution with initial wavelength $\lambda_0=20H_0^{-1}$ and initial density contrast $\delta_0=0.0005$ in three different base resolutions, namely $N_{LR}=128$, $N_{MR}=144$, $N_{HR}=160$. 
Figure~\ref{HamConvergence}, we show the volume averaged Hamiltonian constraint violation 
\begin{equation}
    \mathcal{H}=R^3+K^2-K_{ij}K^{ij}-16\pi \rho
\end{equation}
and the differences between the base resolutions, indicating second order convergence prior to the black hole formation according to Eq.~\eqref{converge}. 
\begin{equation}\label{converge}
    \frac{\|{\langle  \mathcal{H}_{MR}\rangle-\langle  \mathcal{H}_{LR}\rangle}\|}{\|{\langle  \mathcal{H}_{HR}\rangle-\langle  \mathcal{H}_{MR}\rangle}\|}\approx\frac{1/N_{MR}^p-1/N_{LR}^p}{1/N_{HR}^p-1/N_{MR}^p}
\end{equation}
where $p$ is the order of convergence~\cite{article}. In Figure~\ref{MassConvergence}, we show the evolution of the masses of the black holes formed tracked by the apparent horizon finder on base grid resolution, indicating convergence is achieved.


\clearpage
\bibliographystyle{JHEP}
\bibliography{main.bib}

\end{document}